\documentclass[11pt,twoside]{article}

\usepackage{asp2006}
\usepackage{epsf}
\usepackage{psfig}
\usepackage{graphicx}
\usepackage{lscape}
\usepackage{natbib}
\usepackage{amsmath}

\begin{document}
\title{The power of monitoring stellar orbits}   
\author{S. Gillessen$^1$, F. Eisenhauer$^1$, H. Bartko$^1$, K. Dodds-Eden$^1$, T.K. Fritz$^1$, O. Pfuhl$^1$, T. Ott$^1$, R. Genzel$^{1,2}$}   
\affil{$^1$ Max-Planck-Institut for extraterrestrial physics, 85748 Garching, Germany}    
\affil{$^2$ Physics Department, University of California, Berkeley, CA 94720, USA}

\begin{abstract} 
The center of the Milky Way hosts a massive black hole. The observational evidence for its existence is overwhelming. The compact radio source Sgr~ÊA* has been associated with a black hole since its discovery. In the last decade, high-resolution, near-infrared measurements of individual stellar orbits in the innermost region of the Galactic Center have shown that at the position of Sgr~A* a highly concentrated mass  of $4\times10^6\,M_\odot$ is located. Assuming that general relativity is correct, the conclusion that Sgr~A* is a massive black hole is inevitable. Without doubt this is the most important application of stellar orbits in the Galactic Center. Here, we discuss the possibilities going beyond the mass measurement offered by monitoring these orbits. They are an extremely useful tool for many scientific questions, such as a geometric distance estimate to the Galactic Center or the puzzle, how these stars reached their current orbits. Future improvements in the instrumentation will open up the route to testing relativistic effects in the gravitational potential of the black hole, allowing to take full advantage of this unique laboratory for celestial mechanics.
\end{abstract}


\section{Observations}
\begin{figure}
\begin{center}
\includegraphics[width=0.8\columnwidth]{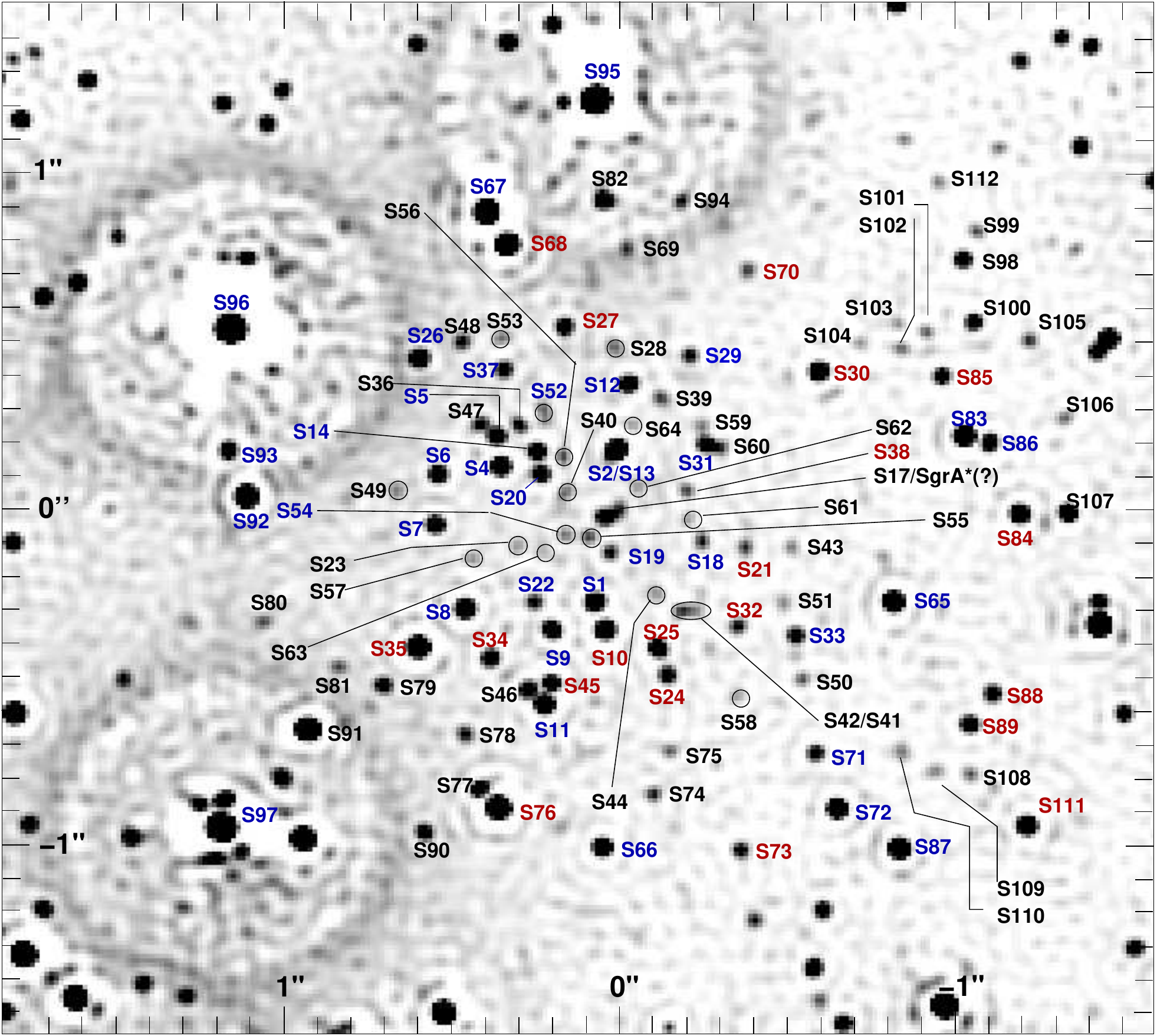}
\caption{\label{fig:findchart} Finding chart of the innermost region of the nuclear star cluster in the Galactic Center. The image is based
on a natural guide star adaptive optics image from NACO at UT4 (Yepun) of the VLT on 2007 July 20 in the H band. Stars that are unambiguously identified in several images have designated names, ranging from S1 to S112. Blue labels indicate early-type stars, red labels late-type stars. Stars with unknown spectral type are labelled in black.}
\end{center}
\end{figure}

Observationally, the Galactic Center (GC) region is characterized by the high extinction screen along the line of sight through the Milky Way plane. In order to observe stars in the GC, one needs to go to the near-infrared (NIR). Given the small angular scales and the crowding of stars, high resolution methods have been the clue for detecting the stellar orbits. In the 1990s NIR observations were using the Speckle technique \citep{Eckart:1996p163,Ghez:1998p118}. Since 2002 adaptive optics observations, both for imaging and spectroscopy are performed routinely, either using natural guide stars or laser guide stars \citep{Schodel:2002p153,Eisenhauer:2005p117,Ghez:2005p1681}. The images obtained at the diffraction limit of the telescope ($\theta_\mathrm{FWHM}\approx 55\,$mas for the K-band around $2.2\,\mu$m at an $8\,$m-telescope) reveal $\approx 100$ stars within $r=1''$ of the massive black hole (MBH, figure~\ref{fig:findchart}).
The brightest of these so called S-stars, S2, has an apparent magnitude of $m_\mathrm{K}=14$. The high density of stars limits how faint a star one can detect in the central arcsecond. Under good conditions, stars down to $m_\mathrm{K}=18.5$ can be detected.
The astrometric precision from such data is in the $200-300\,\mu$as range for bright stars \citep{Ghez:2008p945,Fritz:2010p2374}.
Spectroscopically, the use of adaptive-optics assisted integral field spectroscopy is state of the art. In the central arcsecond, late-type stars as faint as $m_\mathrm{K}\approx17$ can be identfied, given the prominent CO absorption features around $2.3\,\mu$m. For early-type stars, the current limit is $m_\mathrm{K}\approx16.5$, and the detection of the Brackett-$\gamma$ line is hampered by the presence of line emission from the background gas. Radial velocities can be measured with a precision of $\approx15\,$km/s for the brightest early-type stars and for many late-type stars, limited by the accuracy of the wavelength calibration. For fainter early-type stars, the errors are dominated by the signal-to-noise ratio and typical errors are between $50-100\,$km/s.
From these astrometric and spectroscopic measurements the orbits of $\approx30$ stars are known \citep{Ghez:2005p208,Eisenhauer:2005p117,Gillessen:2009p1117}. Surprisingly, a large fraction of these stars deep in the potential of the MBH are young. Their existence in the place where they are observed poses a puzzle \citep{Ghez:2003p178}. Due to the tidal forces of the MBH they cannot have formed in-situ, but their age is much smaller than the migration time scale.

\section{The orbit of S2}
\begin{figure}
\begin{center}
\includegraphics[width=0.32\columnwidth]{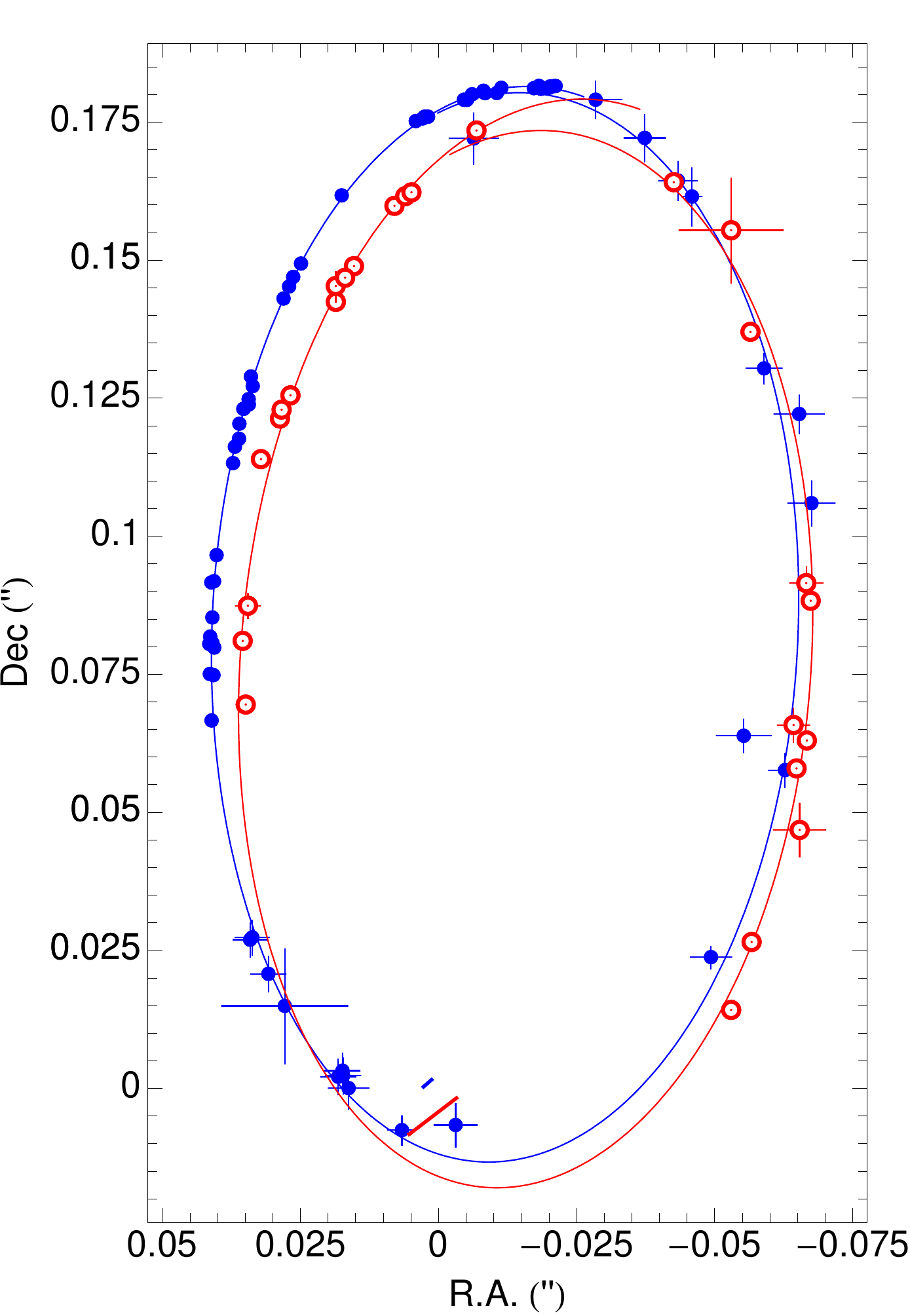}
\includegraphics[width=0.32\columnwidth]{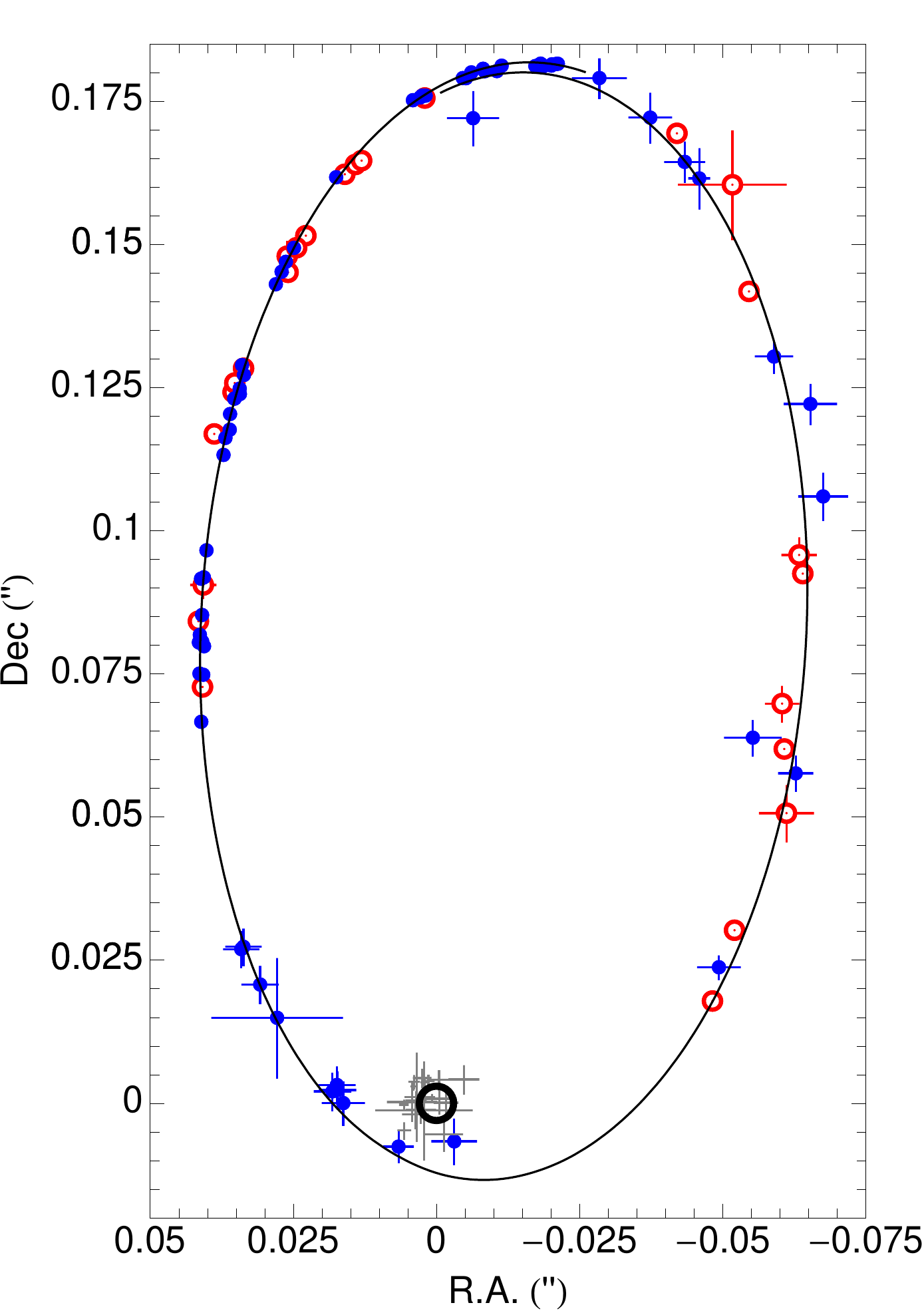}
\includegraphics[width=0.32\columnwidth]{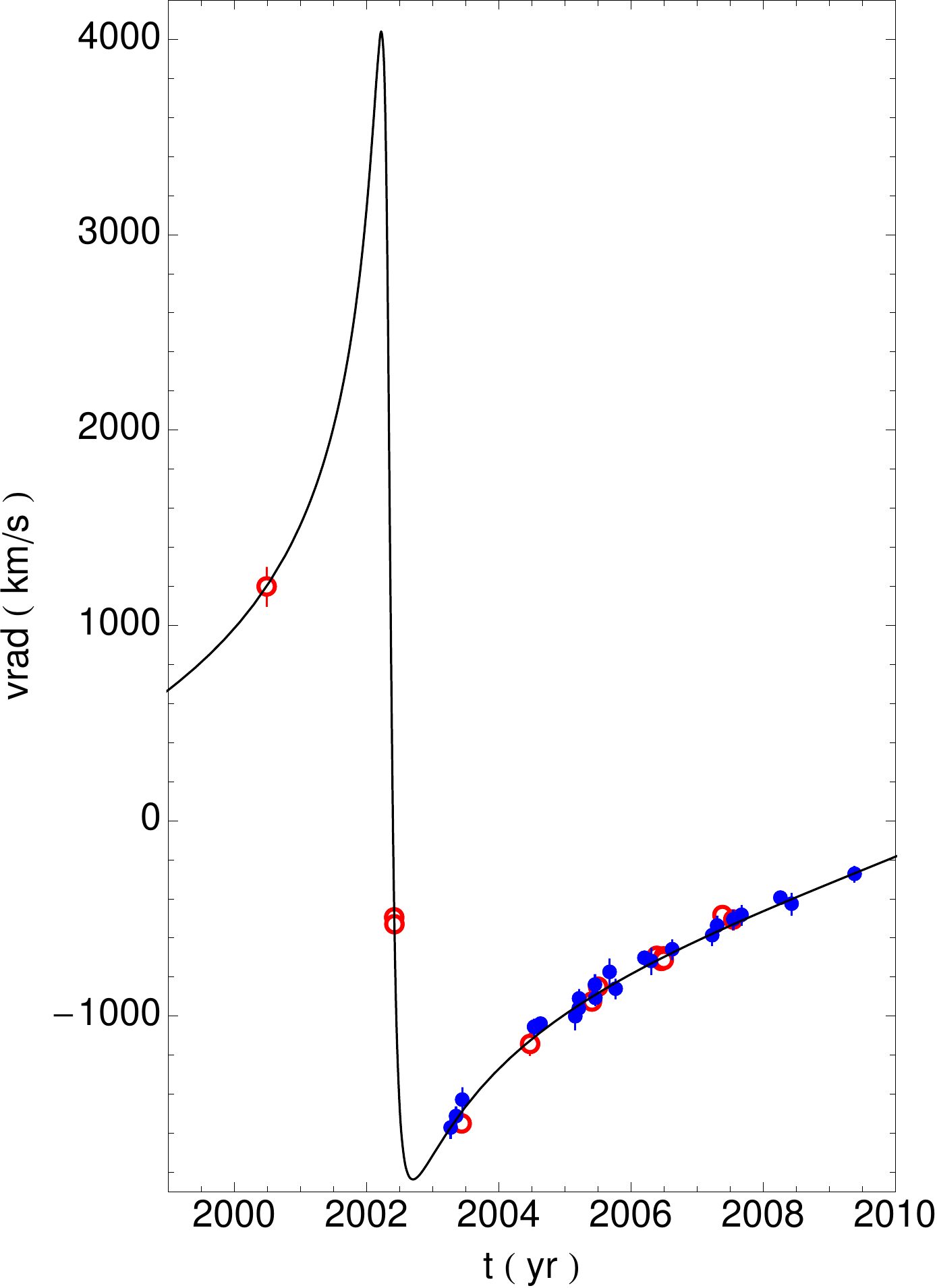}
\caption{\label{fig:s2orbit} The orbit of S2. Left: \citet{Ghez:2008p945} (red, open symbols) and \citet{Gillessen:2009p1117} (blue, filled symbols) have published orbital data for S2. Simply overplotting the two data sets and orbit fits reveals an apparent mismatch larger than the precision of the individual measurements. Middle: \citet{Gillessen:2009p2382} showed that the difference can be understood completely by the uncertainties of the two coordinate systems. A simple offset in position and velocity is sufficient to make the two data sets agree. Right: The radial velocity data.}
\end{center}
\end{figure}

Since the beginning of the observations in 1992, S2 has completed a full revolution around Sgr~A* (figure~\ref{fig:s2orbit}). Its orbital period is $15.9\,$years, the semi-major axis of the orbit is $125\,$mas \citep{Ghez:2008p945,Gillessen:2009p1117,Gillessen:2009p2382}. Since the proper motion is measured in angular units per time, but the radial velocity in physical length units per time, an orbit model contains the distance to the system as parameter. Hence, S2 allows for a geometric determination of $R_0$, the distance to the GC \citep{Salim:1999p1362}. The current best such estimate \citep{Gillessen:2009p2382} is
\begin{equation}
R_0 = 8.28 \pm 0.15|_\mathrm{stat} \pm 0.30|_\mathrm{sys} \, \mathrm{kpc}\,\,.
\end{equation}
Its accuracy is limited by systematic uncertainties. The corresponding value for the central mass is
\begin{equation}
M= 4.30 \pm 0.06|_\mathrm{stat} \pm 0.35|_\mathrm{R_0} \times 10^6 \,M_\odot\,\,.
\end{equation}
The first error gives the uncertainty if the value of $R_0$ were known perfectly well, the second, dominant term is the error induced by the uncertainties on $R_0$. This indicates that the $M$ and $R_0$ are strongly correlated, empirically we find $M\propto R_0^{\,2}$. The exponent here would be 3 for a purely astrometric data set and 0 for a purely spectroscopic one.

The systematic uncertainties had been underestimated in previous works \citep{Eisenhauer:2003p128}. A large part of them comes from the fact that it is hard to establish an absolute reference frame in the NIR in the GC region \citep{Reid:2007p169}. This is illustrated by the fact, that the published data sets do not agree at first glance (figure~\ref{fig:s2orbit}). However, allowing for an offset in position and velocity between the two coordinate systems suffices to bring the measurements into agreement \citep{Gillessen:2009p2382}.

The NIR coordinate system is relying on the astrometric measurements of SiO maser stars that are visible both in the NIR and in the radio regime \citep{Reid:2007p169}. By construction, the radio source should be at rest at the origin of the NIR coordinate system. The position of the central mass observationally is a distinct quantity and thus it is worth to compare how well the positions of mass and Sgr~A* do agree. We find that they coincide to within less than $3\,$mas.

\citet{Mouawad:2005p191}, \citet{Ghez:2008p945} and \citet{Gillessen:2009p1117} have tested, how much mass inside the S2 orbit can be in an extended configuration. These studies consistently find that at most $\approx 5$\% of the mass of the MBH can reside at radii between the peri- and the apocenter distance of the S2 orbit. For example, \citet{Gillessen:2009p1117} find that the fraction $\eta$ of the extended mass to the MBH mass is
\begin{equation}
\eta = 0.018 \pm 0.014|_\mathrm{stat} \pm 0.005|_\mathrm{model} \,\,.
\end{equation}
 
The measurement uncertainty is still too large to give astrophysically relevant constraints. Various arguments point towards a value of $\eta\approx10^{-3}-10^{-4}$.
\begin{itemize}
\item The drain limit gives the maximum number of objects that in a steady state situation can reside close to a MBH. From the work of \citet{Alexander:2004p168} follows that if the dark mass consists of stellar black holes with $10\,M_\odot$, one would expect $\eta \leq 1.1\times 10^{-3}$.
\item Theoretical modelling, either in a Fokker-Planck approximation or using explicit N-body simulations yield consistently $\eta\approx 5\times10^{-4}$ \citep{Hopman:2006p155,Freitag:2006p1780}.
\item The overdensity of X-ray binaries in the central parsec \citep{Muno:2005p137} also estimates the number of stellar black holes in the GC. The measured density corresponds to $\eta\approx2\times10^{-4}$ \citep{Deegan:2007p146}.
\end{itemize}
The amount of cosmological dark matter is completely negligible compared to the expected astrophysical background density. The dark matter density is at least two order of magnitudes smaller \citep{Vasiliev:2008p1092}.

The only hypothesis which can be tested in the near future is, whether the S-stars have reached their orbits via normal two-body relaxation from a birth place in the region of the disks of young, massive stars \citep{Paumard:2006p162,Lu:2009p1698,Bartko:2009p1977}. This would require significantly more mass than what is seen in stars. Stellar black holes could lower the relaxation time such that the lifetime of a B-star would suffice to migrate from the disk region into the central arcsecond. The corresponding value of $\eta$ is $\approx 0.03$, relatively close to what can be measured. However, such a scenario is disfavored anyhow.

The prospects for detecting relativistic effects are much better. \citet{Zucker:2006p194} showed that the relativistic effects of order $\beta^2$ in the radial velocity are detectable in the orbit of S2 with today's technique during a pericenter passage. Besides the classical Romer effect, these are the transverse Doppler effect and gravitational redshift. Here, it is important to realize that it is not sufficient to compare for a given fixed orbit the Keplerian and relativistic fits. Instead, one needs to compare the two fits for a given set of data, since the orbital elements need to be determined from the same data. The relativistic effects are degenerate to a strong degree with the orbital elements when fitting orbits.

The largest effect detectable in astrometry is the prograde pericenter precession due to the Schwarzschild correction of the Newtonian potential. For S2, the difference between best fitting Keplerian and relativistic orbit is smaller than $0.5\,$mas at all times for the next years, and it will be considerably more difficult to detect the precession than the effects on the radial velocity. This can also be seen from precision by which the position angle of the ascending node is determined. Currently, after 18 years of monitoring, $\Delta \omega=0.84^\circ$, while the precession
amounts to $0.22^\circ$ per revolution \citep{Rubilar:2001p186}.

\section{The limits of astrometry}

\begin{figure}
\begin{center}
\includegraphics[width=0.8\columnwidth]{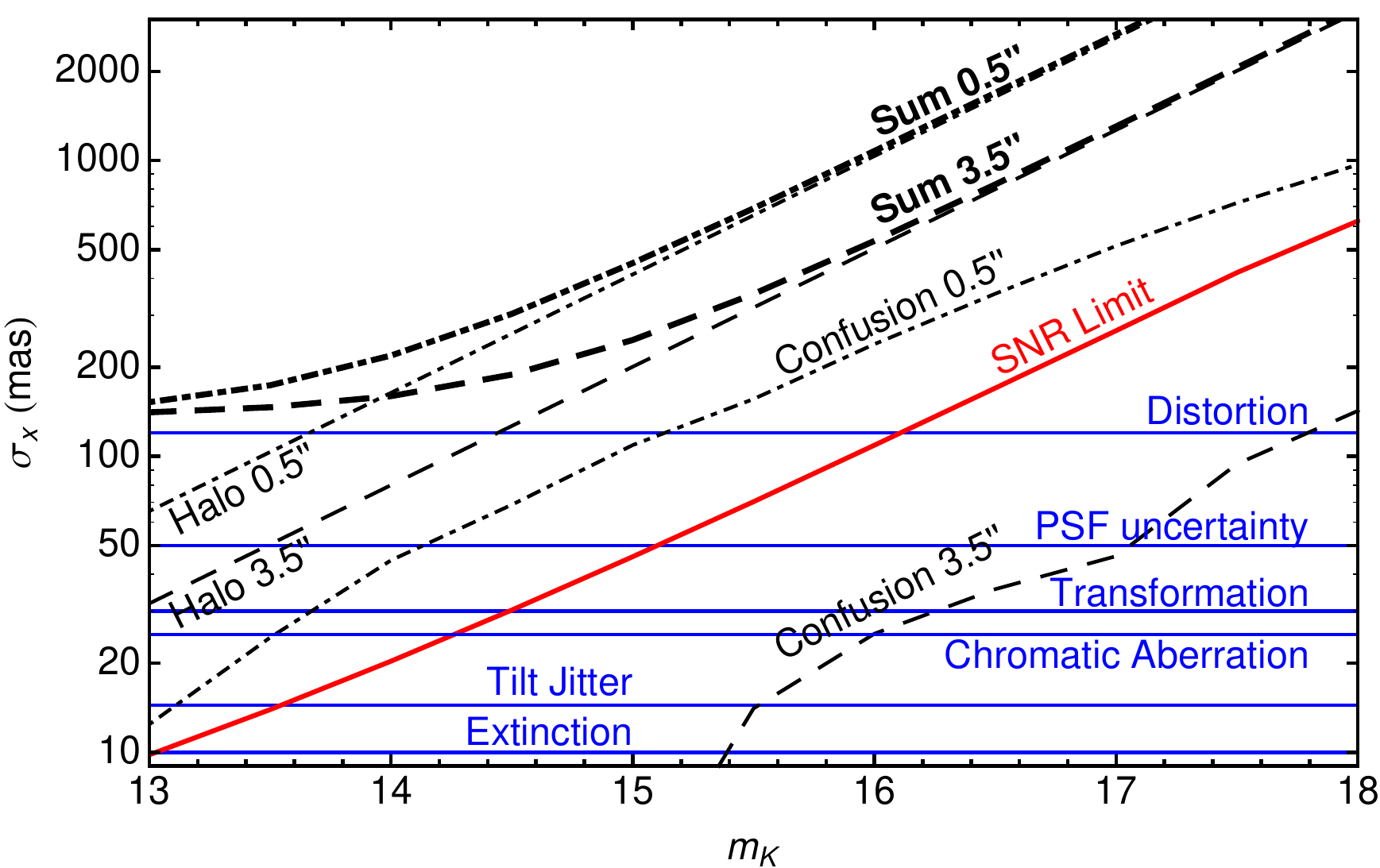}
\caption{\label{fig:limits} Contributions of various effects to the errors of astrometric positions as a function of brightness of the star considered. }
\end{center}
\end{figure}

\citet{Fritz:2010p2374} have investigated the limits of adaptive-optics based astrometry in the crowded GC region. The results are summarized in figure~\ref{fig:limits}. For the brightest S-stars, the limitation currently is the knowledge of the optical distortions of the camera. For fainter stars, the uncertainty is dominated by the so-called halo noise. The term refers to the fact that the imperfectly determined seeing halos of the brighter stars perturb the positions of the fainter stars. The effect depends on the level of crowding (and thus on distance to Sgr~A*) and most importantly on the Strehl ratio. The signal-to-noise ratio is not the limiting factor. Obviously, higher Strehl ratios would improve the astrometric precision, and the same is true if one would be able to increase the spatial resolution.

The next big step in angular resolution is expected to come from NIR interferometry. The resolution which the VLT interferometer can achieve is as good as $3\,$mas in the the K-band. Due to the baselines of $\approx100\,$m length, this exceeds even the performance of the next generation of extremely large telescopes. At that resolution, one should be able to observe stars within the central $100\,$mas that would have orbital periods of $\approx1\,$year. The relativistic precession would amount to several degrees per revolution and should be easily detectable within few years of monitoring. The number of stars expected from extrapolating the luminosity function \citep{Bartko:2010p2407} and the radial density profile \citep{Genzel:2003p151,Schodel:2007p144} in that radial range is between three and ten for magnitudes between $m_\mathrm{K}=17$ and 19.

\section{Statistical properties of orbital elements}

\begin{figure}
\begin{center}
\includegraphics[width=0.75\columnwidth]{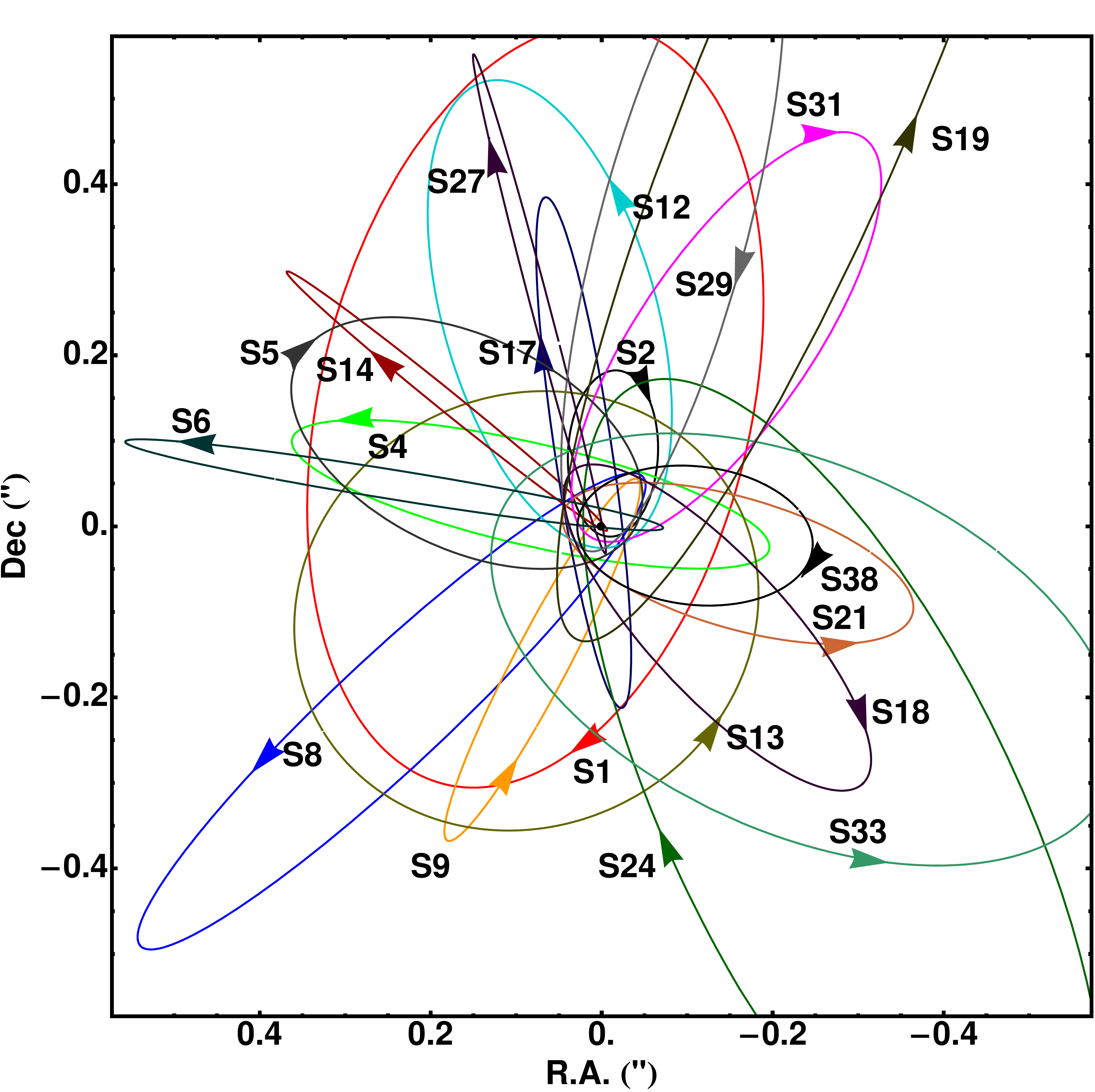}
\caption{\label{fig:orbits} Illustration of the stellar orbits determined so far. Here we show the innermost 20 of the known orbits.}
\end{center}
\end{figure}

\begin{figure}
\begin{center}
\includegraphics[width=0.85\columnwidth]{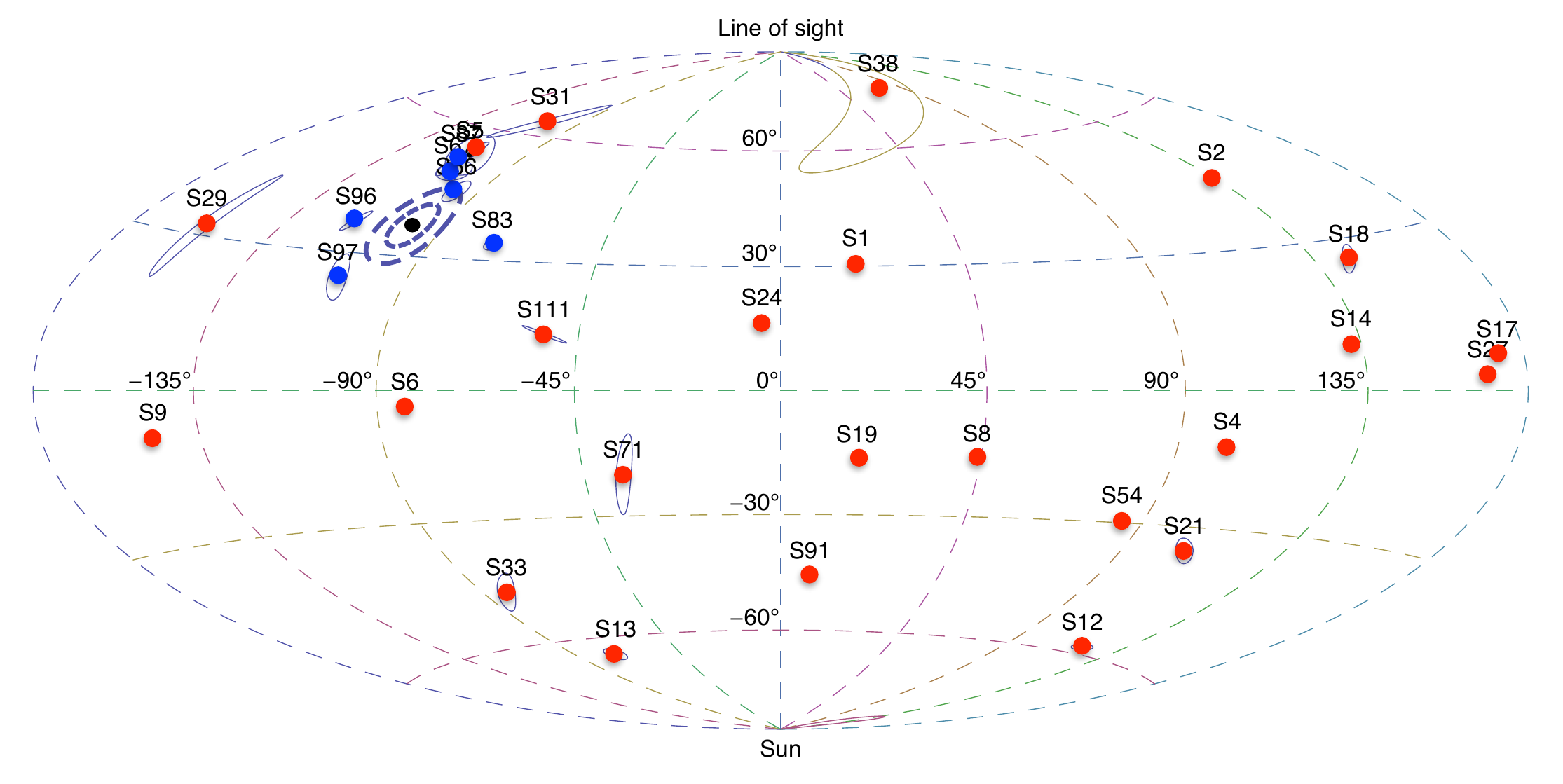}
\caption{\label{fig:angmom} Orientations of the orbital planes for the known orbits in the GC. The blue dots represent stars that belong to the clockwise rotating stellar disk. The red dots are the other stars, which appear to be randomly distributed.}
\end{center}
\end{figure}

Given that $\approx30$ orbits have been determined so far (figure~\ref{fig:orbits}), one can start to look into the distributions of orbital elements \citep{Gillessen:2009p1117}. They are extremely valuable clues for the enigmatic formation of the S-stars.

The most striking feature in the distribution of orbital elements is the distribution of orbital planes. While for the central S-stars the distribution is consistent with a random distribution, we find a group of six orbits at larger radii ($r\approx1''$) that share a common plane. This explicitly confirms the existence of the clockwise stellar disk \citep{Paumard:2006p162,Lu:2009p1698,Bartko:2009p1977} that previously was defined only in a statistical sense. The six orbits all have a moderate eccentricity ($e\approx0.4$) and the brightness of these stars is larger than for the stars at smaller radii. We indeed see two populations, the disk stars (more massive, orbiting in a common plane, moderate eccentricity) and the B-type S-stars (randomly oriented orbits with higher eccentricity).

Secondly, the distribution of eccentricities (of non-disk stars) is slightly hotter than for a thermal ($n(e)\propto e$) distribution:
\begin{equation}
n(e) \propto e^{2.8\pm0.8}
\end{equation}
This potentially is a key to the formation of the S-stars. In the Hills mechanism \citep{Hills:1991p1802} scenario, massive perturbers (for example molecular clouds) scatter field stars into a near loss-cone orbit \citep{Perets:2007p560}. If that happens to a binary, the three-body interaction with the MBH at closest approach can break the binary, leaving one member as S-star and ejecting the other star at high velocity, which might explain the existence of hypervelocity stars \citep{Brown:2005p1103,Brown:2006p2605}. Such a scenario indeed would yield an eccentricity distribution hotter than thermal, however it would be much hotter \citep{Ginsburg:2006p2591,Alexander:2007p493}. Hence an additional relaxation mechanism needs to be invoked, maybe resonant relaxation \citep{Hopman:2006p155}. Clearly, determining the eccentricity distribution more accurately is important to the  binary-capture scenario.

The distribution of semi-major axes tests the density profile: $n(r)\propto r^\alpha$ for $n(a)\propto a^{\alpha+2}$ \citep{Schodel:2003p180}. Our measurements for $r<0.5''$ yield $n(r)\propto r^{-1.1}$, which agrees well with previous works at larger radii \citep{Genzel:2003p151,Schodel:2007p144}.

\section{Constraining the presence of an intermediate mass black hole}

An intermediate mass black hole (IMBH) in the GC is proposed as alternative to the binary capture mechanism \citep{Merritt:2009p2457,Gualandris:2009p2530}. Also, an IMBH might play an important role for the young, massive stars. If they have reached the GC as part of an inspiraling cluster, the IMBH might have helped to bind the cluster such that it was able reach distances of a few arcseconds. \citep{Hansen:2003p152,Gurkan:2005p238}.

The current limits on the motion of Sgr~A* obtained from stellar orbits are not competitive with the radio measurements of the motion of Sgr~A* \citep{Reid:2004p190,Merritt:2009p2457} since they only exclude a region in the parameter space of mass and distance of the IMBH at relatively small distances and large masses that is excluded form the radio data already.

\citet{Harfst:2008p1880} showed than an IMBH would change the orbital elements of the S-stars continuously, and maybe by an amount which is not much below the current measurement accuracy. Hence, testing the orbital elements for constancy can probe the presence of an IMBH. In a more extreme form, one might also hope to be lucky to observe a gravitational encounter between an IMBH and an S-star. 

\section{The nature of the TeV source in the Galactic Center}

\begin{figure}
\begin{center}
\includegraphics[width=0.80\columnwidth]{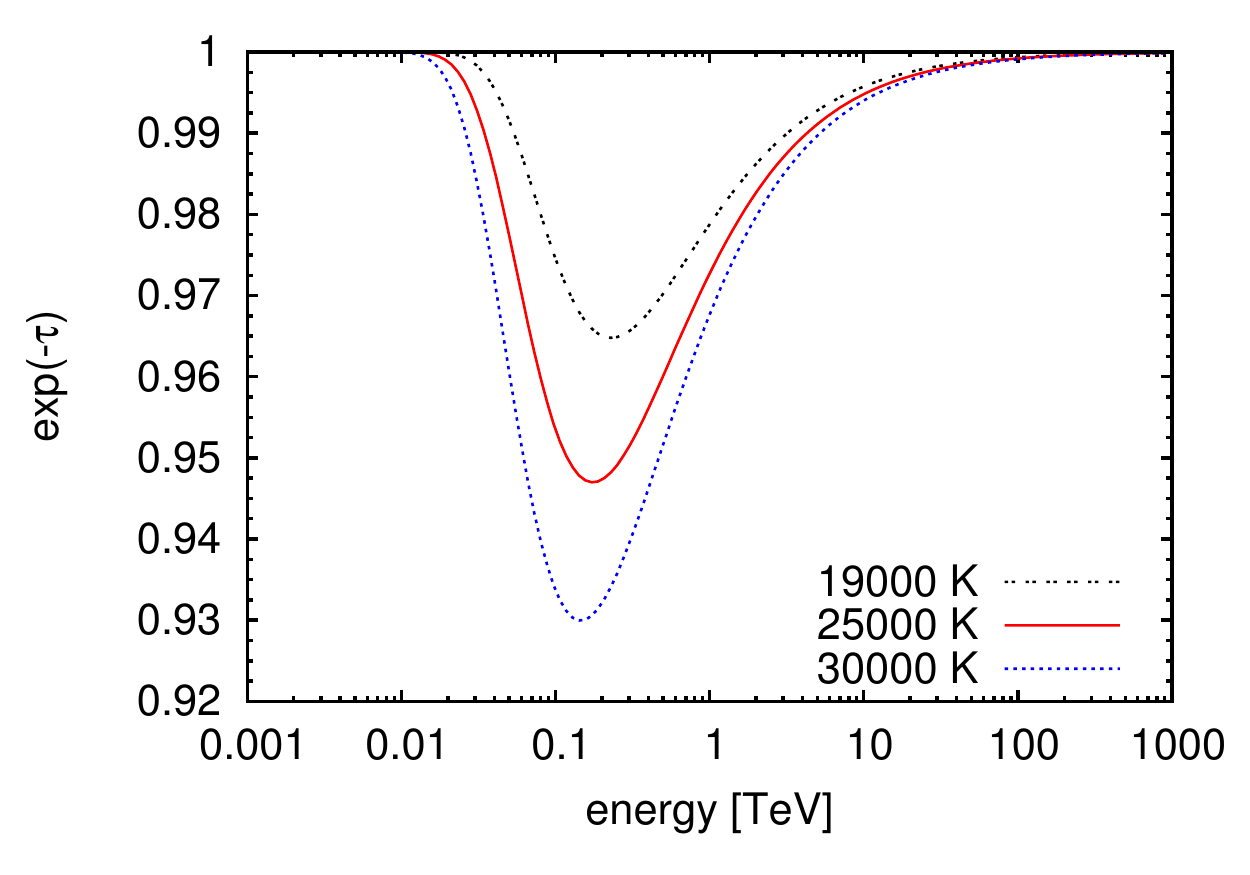}
\caption{\label{fig:tev}. Energy dependence of the absorption of the TeV radiation from Sgr~A* due to a star with given temperature for an angular separation of $1\,$mas. From \citet{Abramowski:2009p2545}.}
\end{center}
\end{figure}

At the position of Sgr~A* also a TeV source is observed \citep{Aharonian:2004p1219}. It is unlikely that the emission is due to dark matter annihilation \citep{Horns:2005p2540}. Instead, it probably is associated either with Sgr~A* itself or a pulsar wind nebula a few arcseconds north-west of Sgr~A* \citep{Acero:2009p2543}. The pointing uncertainties of the Cerenkov telecopes make it unlikely that from the direct measurements one will be able to decide which source is the origin for the TeV signal from the GC. 

The S-stars offer a neat probability to test, whether Sgr~A* is the TeV source. The (UV) radiation field around a star is opaque to TeV gamma rays. Hence, if an S-star crosses the line of sight to Sgr~A*, one might observe a decrease of the TeV flux. \citet{Abramowski:2009p2545} have calculated the size of the effect for the known S-stars (figure~\ref{fig:tev}). A flux decrease of a few percent over the course of a few months would be the typical signal of such an pair production eclipse. Future TeV detectors should be able to detect this. The absence of the signal in turn would exclude that Sgr~A* is the TeV source.



\bibliographystyle{asp2010}
\bibliography{gillessen}

\end{document}